\documentclass[3p,times,twocolumn]{article}

\usepackage{amssymb}

\usepackage[figuresright]{rotating}

\begin{document}

\begin{titlepage}
\hskip 11 cm { \bf IFJ PAN-IV-2014-18}

\vskip 1 cm

\centerline{ \Large\bf The $\tau$ leptons theory and  experimental data: 
}
\centerline{ \Large\bf Monte Carlo, fits, software and systematic 
errors$^{\dagger,\dagger\dagger}$. }

\vskip 1 cm

\centerline{\large \bf Z. Was$^*$}

\vskip 1 cm
\centerline{\bf IFJ PAN, PL-31342 Krakow, Poland}

\vskip 2 cm

\begin{abstract}

Status of    
  $\tau$ lepton decay Monte Carlo generator {\tt TAUOLA} is reviewed. 
Recent efforts on development of new hadronic currents are presented.
Multitude new channels for anomalous $\tau$ decay modes and parametrization 
based on defaults used by BaBar collaboration are introduced. Also
parametrization based on theoretical considerations are presented as an alternative. 
 Lesson from  
comparison and fits to the BaBar and Belle
 data is recalled. It was found that as in the past, 
in particular at a time of comparisons with  CLEO and ALEPH data,
proper fitting, to as detailed as possible representation of the 
experimental data, is essential for appropriate developments of 
models of $\tau$ decays.

In the later part of the presentation,
use of the {\tt TAUOLA} program for phenomenology of $W,Z,H$ decays at LHC 
is adressed.
Some new results, relevant  for QED bremsstrahlung in such decays are presented as well. 


\end{abstract}



\vfill

\noindent
{ \bf IFJ PAN-IV-2014-18, \\ December 2014}

\vskip 13 mm
\footnoterule
{\footnotesize
\noindent
{$^\dagger$Presented on the 13th International Workshop on Tau Lepton Physics
Aachen, Germany, 15-19 September, 2014.}

\noindent
{$^{\dagger\dagger}$This research was supported in part by
the Research Executive Agency (REA) of the European Union under
the Grant Agreement PITN-GA-2012-316704 (HiggsTools) and by funds of Polish National Science
Centre under decision DEC-2011/03/B/ST2/00107.}

\noindent
$^*$
{\it Email address:} z.was@cern.ch,
{\it URL:} http://wasm.web.cern.ch/wasm/

}
\end{titlepage}


\section{Introduction}
\label{intro}

The {\tt TAUOLA} package
\cite{Jadach:1990mz,Jezabek:1991qp,Jadach:1993hs,Golonka:2003xt} for simulation
of $\tau$-lepton decays and
{\tt PHOTOS} \cite{Barberio:1990ms,Barberio:1994qi,Golonka:2005pn} for simulation of QED radiative corrections
in decays, are computing
projects with a rather long history. Written and maintained by
well-defined (main) authors, they nonetheless migrated into a wide range
of applications where they became ingredients of
complicated simulation chains. As a consequence, a large number of
different versions are presently in use. Those modifications, especially in case of
{\tt TAUOLA}, are   valuable from the physics point of view, even though they
 often did not find the place in the distributed versions of
the program.
From the algorithmic point of view, versions may
differ only in  details, but they incorporate many specific results from distinct
$\tau$-lepton measurements or phenomenological projects.
Such versions were mainly maintained (and will remain so)
by the experiments taking precision data on $\tau$ leptons.
Interesting from the physics point of view changes are still
developed in {\tt FORTRAN}.
That is why, for convenience of such partners, part of the
{\tt TAUOLA} should remain in {\tt FORTRAN} for a few forthcoming years.

The program structure did not change significantly 
since previous
$\tau$ conference \cite{Was:2012ne}, nowadays however, the C++ 
implementation become dominant for many aspects of  the project.
In the following, we will  concentrate on physics extensions and novel applications.
We will stress importance of the three aspects of the work:
(i)  construction and implementation of hadronic currents for $\tau$ decay 
currents obtained from models (evaluated from QCD) (ii) presentation of 
experimental data in a form suitable to fits (iii) preparations of algorithms 
and definition of distributions  useful for fits.

We have prepared two
new set of currents, the first based mainly on theoretical consideration, 
the second on an effort of BaBar collaboration. They are ready to be 
integrated into main distribution tar-balls for {\tt FORTRAN} and {\tt C++} 
applications.
Further work in evaluation if such parametrizations are better suitable for physics purposes is on-going. Weighted event 
techniques useful  for fits were studied as well.
Analyses of high precision,
high-statistics  data from Belle and BaBar may find them useful, similar as LHC experiments. 

Our presentation is organized as follows:
Section 2  is devoted to presentation of  optional initialization for 
 {\tt TAUOLA} which is based on the program version evolving in BaBar 
starting from a variant presented on $\tau$ conference of 2004 \cite{Was:2004dg}.
This version is supplemented with multitude of anomalous $\tau$ decay modes
as well as with parametization of our theoretical works of the last decay 
(of different level of theoretical sophistication depending on decay channel). Possibility to replace 
hadronic currents or matrix elements with the user provided C++ code 
is introduced. 
In Section 3 we concentrate on   {\tt PHOTOS} Monte Carlo for
radiative corrections in decays. The new version of the program is now 100 \%
in C++.
Section 4 is for the interfaces of {\tt TAUOLA} and {\tt PHOTOS} based
on  {\tt HepMC} and written in {\tt C++}. Work on interface to
genuine weak corrections, transverse spin effects and new tests
and implementation
bremsstrahlung kernels is presented as well. Next, the algorithm of  {\tt TauSpinner} is presented. It calculates
 weights to manipulate properties of 
the event sample accordingly to changed 
assumptions for the hard process dynamic, or due to changed level of 
implementation of spin effects.
 Summary Section 5, closes the presentation.

Because of the limited space of the contribution,
 some results  will not be presented in the
proceedings. They find their place in
publications, prepared with co-authors listed in the References.
For these works,  the present paper may serve as an advertisement.

\section{ New currents in  {\tt TAUOLA} Monte Carlo}

In other talks \cite{Shekhovtsova:2014xsa,Roig:2014kwa} of the conference, it was shown how  Resonance Chiral Lagrangian
approach was used for calculations of  hadronic currents
to be installed in  {\tt TAUOLA}. We do not need to repeat it here. 
In~\cite{Shekhovtsova:2014xsa} it was  stressed that
 details, such as additional  resonances, more specifically the
$f_2(1270)$, $f_0(1370)$ and $a_1(1640)$, observed  by CLEO long time ago \cite{Asner:1999kj} can not be introduced if fits to  one-dimensional 
invariant mass spectra of two- and three-pions systems are  
only used. In Ref.~ \cite{Asner:1999kj}  as an input for  
 parametrization of {\tt TAUOLA} currents ({\tt cleo} 
parametrization\footnote{Note that for this parametrization, 
differences between hadronic currents of $\tau \to \pi^+\pi^-\pi^- \nu$ 
and $\tau \to \pi^-\pi^0\pi^0 \nu$   were ignored; 
isospin symmetry was imposed ($\rho\pi$ dominance). Version of the current without this constraint
is nonetheless distributed with {\tt TAUOLA} (all versions), 
but as an non-active option. 
On the other hand, it  not only was developed and used by CLEO, but existed
(as one of  options) in  BaBar software. I am thankful to Swagato Banerjee
for clarification.}
\cite{Golonka:2000iu}), 
two-dimensional mass scattergrams were used. This should be considered as a 
minimum for the comparisons with the present day data as well. In fact, 
already CLEO used more detailed 
representation of the data in \cite{Browder:1999fr}. 
It may be of interest to repeat such data analysis, with the help of observables,
as the ones presented in \cite{Kuhn:1992nz}, but
adopted to the case of relativistic tau-pair production of Belle or BaBar
experiments. 

Physics of $\tau$ lepton decays requires sophisticated strategies for the
confrontation of phenomenological models with experimental data. On one hand,
high-statistics experimental samples are collected, and the obtained precision is
high, on the other hand, there is a significant cross-contamination between distinct
$\tau$ decay channels. Starting from  a certain precision  level all channels
need to be analyzed simultaneously. Change of parameterization for one channel
contributing  to the background to another one may be important for the fit of
its currents. This situation leads to a complex configuration where a multitude of parameters (and models)
needs to be simultaneously confronted with a multitude of observables.
One has to keep in mind that the models used to obtain distributions in
 the fits may require refinements or even substantial rebuildings as a consequence
of comparison with the data. The topic was covered in detail in the $\tau$ Section 
of Ref.~\cite{Actis:2010gg}. At present our comparison with the data still does not 
profit from such  methods.

Let us give some details. We may want 
to  calculate for each generated event (separately for 
decay of $\tau^+$ and/or $\tau^-$) alternative weights; the ratios
of the matrix element squared obtained with new currents,
and the one actually used in generation. Then, the vector of weights can be obtained
and used in fits.
We have checked that such a solution not only can be easily installed into
{\tt TAUOLA} as a stand-alone generator, but it can also be incorporated into
the simulation frameworks of Belle and BaBar collaborations. For practical reasons 
use of semi analytical distributions is much easier. It enables much faster 
calculation of errors for fit parameters including correlations, but experimental distributions must be available in unfolded form.
This was an important ingredient of our work for fits of 3$\pi$ currents 
obtained in \cite{Shekhovtsova:2012ra}. We have found that modifications of 
the currents were necessary to obtain results given in \cite{Nugent:2013hxa}. 
It is not clear, if such fitting, without additional help of observables as 
in \cite{Kuhn:1992nz} can be used for the 
$KK\pi\nu_\tau$ and $K\pi\pi\nu_\tau$ $\tau$ decay channel, even if two-dimensional 
scattergams are available. One has to keep in mind that if experimental data 
are available as one or at most two dimensional histograms then resulting currents 
rely on the models. With the present day precision of the data, even an approach of 
Resonance Chiral Lagrangian should not be expected to have sufficient 
predictive power to describe multidimensional distributions from the constraints of 
fits to one or two dimensional histograms. This limitation is clearly visible in
results for $4-\pi$'s currents of ref.\cite{Bondar:2002mw}.

Keeping all these limitations in mind, we have nonetheless prepared
 currents for 
{\tt TAUOLA} based on
Refs.~\cite{Nugent:2013hxa,Bondar:2002mw,Fujikawa:2008ma,Kuhn:2006nw}, 
respectively for 3, 4,  2  and 5 $\pi$'s final states.
This is  now available  (upon 
individual requests) for {\tt FORTRAN} and {\tt C++} users,
together with further possibility that user introduces  own  C++ currents and together
with  BaBar initialization of currents.

\section{{\tt PHOTOS} Monte Carlo for bremsstrahlung; \newline its systematic uncertainties}
\def\CCol{{\tt SANC}}
Over the last two years no major upgrades for functionalities 
were introduced into {\tt PHOTOS} Monte Carlo. On the other hand, technical 
changes were introduced.
In the last couple of months, for the version of the code
prepared for the C++ environment~\cite{Davidson:2010ew}, transition to C++
was completed.

Also work on numerical tests and new applications, especially in domain of LHC 
applications was completed. The results are collected in Refs.~\cite{Arbuzov:2012dx,Doan:2013qqa,Jadach:2013aha}. A precision up to sub-permille level
was confirmed.

\section{  {\tt TAUOLA universal interface} and {\tt TauSpinner}  }

In the development of packages such as {\tt TAUOLA} or {\tt PHOTOS}, questions
of tests and appropriate relations to users' applications are essential for
their
usefulness. In fact, user applications may be much larger in size and
human efforts than the programs discussed here.
Good example of such `user applications' are complete environments to simulate
physics process and control detector response at the same time.
Distributions of final state particles are not always of direct interest.
Often properties of intermediate states, such as a spin state of $\tau$-lepton,
coupling constants or masses of intermediate heavy particles are
of prime interest.
As a consequence, it is useful that such intermediate state properties are
under direct control of the experimental user and can be manipulated
to understand detector responses.

In that perspective, an algorithm of {\tt TauSpinner} \cite{Czyczula:2012ny} to study detector response to spin effects
in $Z, W$ and $H$ decays, represents an important development. The program is calculating weights corresponding to changes of the physics assumption. As an input, events stored on the data file are used. At present,
the program can calculate: spin correlation, production matrix element and decay matrix element weights from the kinematical configurations for events previously generated
and stored on the datafile. In this way detector response to variants 
of physics model used in the production process can be studied. Following
original publication,
the {\tt TauSpinner} was
first enriched~\cite{Banerjee:2012ez} with the option to study effects of new physics, such as
effects of spin-2 states in $\tau^+\tau^-$ pairs produced at LHC.
This work was later continued, and in~\cite{Przedzinski:2014pla} an option to study 
transverse spin effects, important for the Higgs parity measurements was developed.
Work on systematic tests was being performed as well \cite{Kaczmarska:2014eoa}. 
Electroweak corrections taken from
Refs.~\cite{Andonov:2008ga,Andonov:2004hi} can be  also used with the help of the weights. 
Also, as it was found  \cite{Banerjee:2007is}  to be  the case for
evaluation of systematic errors for the measurement of  $e^+e^- \to \tau^+\tau^-$  BaBar cross section,
improvements in precision of $\tau$-decays as discussed in Section 2, 
may be of importance for LHC observables as well.


All our programs are available
through the LHC Computing Grid ({\tt LCG}) Project. See
{\tt GENSER} webpage, Ref.~\cite{Kirsanov:2008zz}, for details.
 This is the case for {\tt TAUOLA, TauSpinner}   and for
{\tt PHOTOS}    as well.
The {\tt FORTRAN} predecessors are
available in this way too.

\section{Summary and future possibilities}

Versions of the hadronic currents available for the {\tt TAUOLA} library
until now, were all based on old models and experimental data of 90's.
The alternative implementation of   new currents, based on the Resonance Chiral Lagrangian, or other approaches
 is now prepared
and tested for the decay channels to 2, 3, 4 and 5 pions. For each of these groups of decay channels,
different level of theoretical sophistication and quality of comparisons with the 
$\tau$ decay data was used.
 On top of it,
parametrizations used for the default simulations in BaBar collaboration became 
available. The particular advantage of this option is a
multitude of rare and anomalous $\tau$ decay channels. In fact this list was extended 
even further. With the help of C++ interface, user provided hadronic current(s) 
or decay matrix element(s),
can replace, the ones of the library. 

In contrary, methods for detailed  confrontation
with the experimental data are not developing that fast. We still rely on 
comparison of results with one (at most two) dimensional histograms of invariant 
masses for sub-groups of $\tau$ decay products, defined for each decay channel 
separately and unfolded from the background by experiments, prior to comparisons.
Even though it was postulated already in \cite{Asner:1999kj} that it may not be optimal.

The status of
associated projects: {\tt TAUOLA universal interface } and {\tt TauSpinner}
was reviewed. Also new results for 
the high-precision version of  {\tt PHOTOS} for QED radiative corrections in
decays, were referenced. 
All these programs are ready for {\tt C++} applications
thanks to the {\tt HepMC} interfaces.

Presentation of the {\tt TAUOLA} general-purpose {\tt C++} interface
was given. Electroweak corrections can be used
in calculation of complete spin correlations in $Z/\gamma^*$ mediated
processes.  An algorithm for study, with the help of weights calculated from
kinematic of events stored on data files,  detector responses to: spin effects 
and production process variants in
$Z$, $W$ and $H$ decays was shown. The corresponding program {\tt TauSpinner},
is useful, eg. to study Higgs parity sensitive observables at LHC. 

\centerline{\bf Acknowledgement}

The work on {\tt TAUOLA} would not be possible without continuous help an encouragements
from experimental colleagues.
The work of all  co-authors of the papers devoted to  {\tt TAUOLA} development 
was of great importance.
I hope, that it is clearly visible from  my contribution. 












\end{document}